\documentclass{article}
\usepackage{spconf,amsmath,graphicx,hyperref}
\usepackage{algorithm}
\usepackage{algpseudocode}
\usepackage{amssymb}
\usepackage{xcolor}
\usepackage{cases}

\newcommand{\vct}[1]{\boldsymbol{#1}}
\newcommand{\mat}[1]{\mathbf{#1}}

\title{Self-Localization of Distributed Microphone Arrays \\ using One Source and Acoustic Reflections
\vspace{-2mm}
}

\name{Thomas Guffroy${}^{1,2}$, Jean-Daniel Pascal Prieto${}^{1,2}$, Cédric Foy${}^{2}$, Antoine Deleforge${}^{1}$ \thanks{This work was made with the support of the French (ANR) and German (DFG) national
research agencies through project AWESOME
(ANR-25-CE23-1873). Public AI models were used as coding assistants, while the research ideas and the writing of the paper are solely by the authors. \vspace{-2mm}}
\vspace{-2mm}
}
\address{${}^{1}$Inria, IRMA, Université de Strasbourg, CNRS UMR 7501, Strasbourg, France \\
         ${}^{2}$UMRAE, Cerema, Univ. Gustave Eiffel, Strasbourg, France
\vspace{-4mm}
}
\begin{document}
%
\maketitle
\begin{abstract}
We present a method to estimate the relative position and orientation of two non-coplanar microphone arrays using a room impulse response (RIR) from a single source. The approach exploits early acoustic reflections, retrieved by each array in the form of
a 3D image-source point cloud, using a recently-proposed image-source localization technique. The task is then reduced to aligning the two unlabeled point clouds, which is made difficult by their inherent symmetries and by spurious, missing and noisy estimates. To this aim we introduce ARCTIC,
a point-cloud registration algorithm that performs a ranked search over congruent triangles anchored on the true source. Simulated and real experiments show that ARCTIC strongly outperforms two baselines, achieving mean angular and radial self-localization errors below $5^{\circ}$ and 10~cm across 36 real RIRs measured in 6 different rooms.

\end{abstract}
\vspace{-2mm}
\section{Introduction}
\label{sec:intro}
\vspace{-2mm}
Ad-hoc acoustic sensor networks, made of microphones distributed in an environment, hold the promise of improving the quality of video conferencing, hearing aids, or live sound capture under minimal configuration assumptions. To fully benefit from the spatial diversity of such networks, \textit{e.g.}, for sound source localization \cite{guo2026dual} or beamforming \cite{taseska2014informed}, one would ideally like to know the relative positions of sensors. Because physically measuring this on site is often impractical, \textit{self-localizing} the receivers purely based on audio has received significant research interest, as reviewed in \cite{plinge2016acoustic}.

The general idea is to assume that all microphones are observing a shared set of \textit{acoustic events}, \textit{i.e.}, sounds emitted by point sources at different positions in the environment. The geometrical problem is then solved by aligning these events in time and space based on the physics of sound propagation. Existing approaches differ by the number of acoustic events required, and by the temporal and spatial information assumed known. In the least-informed setting, each network node is an individual microphone at an unknown 3D position, receiving sound with an unknown constant time delay
\cite{gaubitch2013auto, el2023localizing, hu2026robust}. The authors of \cite{gaubitch2013auto} show that using the times of arrival (TOAs) of events at microphones, the theoretical minimal numbers of (microphone, event) are $(5,12)$ or $(10,6)$. Since the problem is highly sensitive to TOA estimation errors, more events are typically needed in practice. The works in \cite{gaubitch2013auto, el2023localizing, hu2026robust} respectively use 20, 12 and 10 acoustic events in reported results, using increasingly advanced and robust optimization schemes.

To reduce the number of events, several authors make use of additional assumptions, \textit{e.g.}, time-synchronized devices \cite{crocco2011bilinear, le2016closed, schrammen2019robust, hasegawa2026sequential}, colinear microphones \cite{jager2009automatic}, far-field sources \cite{thrun2005affine, redondi2009geometric}, sources at known positions \cite{wang2019doa, he2023distributed}, closely-spaced devices \cite{raykar2005position} or devices on a 2D plane \cite{hennecke2011towards, wozniak2023reverberation, hasegawa2026sequential}. A setting of particular practical interest is when the nodes consist of \textit{microphone arrays} that are internally time-synchronized, with known internal geometries \cite{redondi2009geometric, valente2010self, wozniak2018passive, schrammen2019robust, hu2020passive}. One is then left with estimating the relative 6-degrees-of-freedom (6DoF) position and orientation, as well as a possible delay, for each array. 
The authors of \cite{schrammen2019robust} show that the minimum required number of (non-colinear) acoustic events is then 3 in the 3D case.
%
While the techniques reviewed so far typically suffer from acoustic reverberation, a few studies propose to use reverberation to the problem's \textit{benefit}. \cite{schwartz2017blind, chetupalli2018linear} make use of spatial correlations in the diffuse sound field, \cite{nowakowski2015microphone} of room modes, and \cite{dokmanic2014localize, salvati2016sound, tegler2022sensor} of early acoustic reflections. Except in \cite{tegler2022sensor}, this strategy allows self-localization from a \textit{single acoustic event}. This is however at the cost of additional assumptions, \textit{e.g.}, known room geometry \cite{salvati2016sound}, limited reflection order \cite{dokmanic2014localize}, or closely-spaced coplanar microphones \cite{nowakowski2015microphone,schwartz2017blind,chetupalli2018linear}.

\begin{figure*}[t!]
  \centering
  \includegraphics[width=0.9\textwidth]{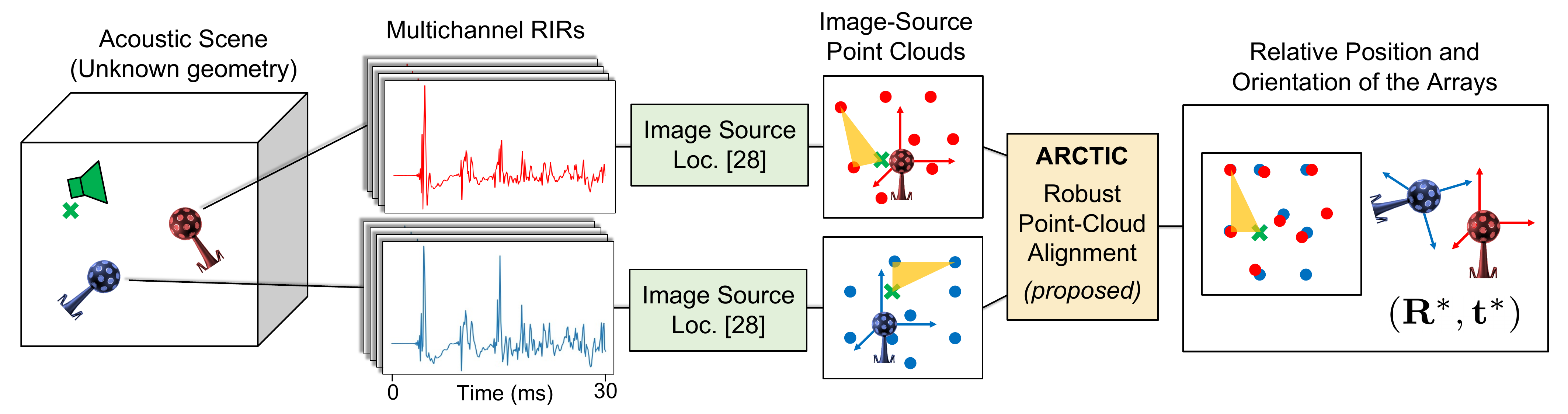}
  \vspace{-4mm}
  \caption{Proposed approach: RIRs are mapped to image-source point clouds which are robustly aligned to self-localize the arrays.
  \vspace{-6mm}}
  \label{fig:proposed_pipeline}
\end{figure*}

In this paper, we consider a minimalist setting that has not been explored in prior literature, to the best of our knowledge. We estimate the relative position and orientation of two synchronized non-coplanar microphone arrays using a room impulse response (RIR) from a single source.
We make use of a recently proposed method that retrieve early acoustic reflections from multichannel RIRs in the form of unlabeled 3D point clouds of \textit{image sources} \cite{prieto2026bridging}. The task is then reduced to aligning the point clouds of the two arrays, which is made difficult by natural symmetries and estimation errors. To tackle this, we introduce \textbf{ARCTIC} for \textbf{A}nchor-based \textbf{R}egistration using \textbf{C}ongruent \textbf{T}r\textbf{I}angle \textbf{C}onsensus.
Experiments 
reveal that the algorithm strongly outperforms two point-cloud registration baselines, unlocking precise and reliable self-localization on both simulated and real data.

\vspace{-2mm}
\section{Background}
\vspace{-2mm}
\subsection{Room Impulse Responses and Image Sources}
\vspace{-2mm}
\label{subsec:RIR_Background}
Given an acoustic point source and a receiver inside a room, the RIR is defined as the time-domain pressure signal at the receiving point assuming the source emits a perfect Dirac impulse. Schematically, the signal is composed of a first peak at the time of arrival of the direct propagation path from source to receiver at the speed of sound, followed by increasingly dense, mixed and attenuated peaks referred to as \textit{echoes}, corresponding to reflected propagation paths attenuated by the room's boundaries and other sound-reflecting objects. 
The early part of RIRs is usually dominated by specular reflections, which can be efficiently modeled by the \textit{image source method} (ISM) \cite{allen1979image}. The ISM asserts that the reflected sound field is equivalent to that obtained by replacing the room's boundaries by iterated mirror attenuated copies of the original sound source across those boundaries, forming a 3D point cloud of \textit{image sources} beyond the room's volume.
Crucially, the theoretical positions of image sources do not depend on the receiver's placement. However, their \textit{visibility} does, and needs to be checked by ensuring that the corresponding reflected path only goes through actual room-boundary faces. The method of this paper relies on two mild assumptions:
    (A1) Microphone arrays are compact enough that their constituting microphones have the same visibility on (early) image sources;
    (A2) The room's shape and placement of the arrays are such that there is a significant overlap between visible image sources by the two arrays. 
%
Note that both assumptions are always verified in the common case of \textit{shoebox} rooms.

In practice, RIRs are measured by emitting a broadband signal such as an exponential sine sweep with a loudspeaker. In turn, the shape of early echoes is strongly impacted by the directive frequency responses of the source and microphones, as can be seen in the examples of Fig.~\ref{fig:proposed_pipeline}.
We recently proposed a two-step approach to localize a 3D image-source point cloud given the early part of a multichannel RIR measured with a compact microphone array, under assumption (A1) \cite{prieto2026bridging}. First, a diffusion-based generative model called \textit{Real2Sim} is used to remove the angular- and frequency-dependent effects of the source, microphones and reflectors from the RIR, yielding a barebone RIR signal made of $\operatorname{sinc}$-like echoes. Second, the model-based 3D gridless image-source localization method from \cite{sprunck22} is used to retrieve the point cloud. In \cite{pascal2026arpege}, the method is shown to accurately localize point clouds of 15 to 40 image sources on real data, with mean angular and radial errors around $5^{\circ}$ and 12~cm.

\vspace{-2mm}
\subsection{Point Cloud Registration}
\vspace{-2mm}
\label{sec:point_cloud_registration}
\textit{Point cloud registration} refers to finding a rigid geometric transformation between two point sets. Let $\mathcal X = \{\vct x_i\}_{i=0}^{N-1}$ and $\mathcal Y = \{\vct y_i\}_{i=0}^{M-1}$ be the two point clouds of image sources extracted using \cite{prieto2026bridging}. Under assumption (A2), retrieving the 6DoF position and orientation of one array with respect to the other is equivalent to finding a rigid transformation $(\mat R, \vct t)$ that \textit{``maximally aligns''} $\mathcal{X}$ and $\mathcal{Y}$, where $\mathbf{R}\in \operatorname{SO}(3)$ is a rotation matrix and $\boldsymbol{t}\in\mathbb{R}^3$ is a translation vector. When point correspondences are known,
this is called a \textit{Procrustes} problem. For this, the \textit{Kabsch algorithm} \cite{kabsch1976solution} provides a method to retrieve the transformation by minimizing the root mean square deviation (RMSD) between the sets, for $M=N\ge3$:
\begin{equation}
    \label{eq:kabsch_rmsd}
    (\mat R^*, \vct t^*) = \operatorname{argmin}_{(\mat R, \vct t)} \textstyle\sum_{i=0}^{N-1} \| (\mat R \vct x_i + \vct t) - \vct y_i \|^2.
\end{equation}
When point correspondences are unknown, they need to be estimated. Iterative methods like \textit{iterative closest point} (ICP) \cite{besl92} or \textit{coherent point drift} (CPD) \cite{myroneko10} alternatively estimate the optimal transformation $(\mat R^*, \vct t^*)$ and the optimal correspondence function $\pi$ until a stopping criterion is reached. Unfortunately, they heavily depend on the initial transformation provided as input and may converge to a local minimum.
Global registration algorithms overcome this, \textit{e.g.}, the 4-point congruent sets (4PCS) method \cite{aiger08}, a \textit{hypothesize-and-verify} algorithm. First, a list of candidate \textit{congruent quadrilaterals} is efficiently built, using geometrical invariants (\textit{hypothesize}). Their congruency is then checked by evaluating the corresponding transformation $(\hat{\mat R}, \hat{\vct t})$ with rigid Procrustes alignment \eqref{eq:kabsch_rmsd}.
Finally, hypothesized transformations are scored (\textit{verify}) by counting how many transformed points land under a prescribed distance $\delta$ to another point in the target set.
%
The transformation with the highest score is retained.

\vspace{-2mm}
\section{Proposed Method: ARCTIC}
\vspace{-2mm}
\begin{algorithm}[t]
\caption{ARCTIC algorithm}
\label{alg:arctic}
\small
\begin{algorithmic}[1]



\Statex \textbf{Input:}
Image-source point clouds $\mathcal{X},\mathcal{Y}\subset\mathbb{R}^3$; anchors $(\vct{x}_0,\vct{y}_0)\in\mathcal{X}\times\mathcal{Y}$; matching threshold $\varepsilon$ (in meters); number of iterations $N_{\mathrm{iter}}$.

\Statex \textbf{Output:}
Matched point pairs $\mathcal{M}^*=\{(\vct{x}_k,\vct{y}_k)\}_{k=1}^K\subset\mathcal{X}\times\mathcal{Y}$; rotation $\mat R^*$; translation $\vct{t}^*$ such that $\vct{y}_k\approx \mat R^*\vct{x}_k + \vct t^*$.

\vspace{1mm}
\State \textbf{Hypothesize} \textit{(Build candidate pairs of anchored triangles)}

    \Statex 
    (a) Build candidate pairs of anchor-originated \textit{edges} $\mathcal{E}$ (Eq. \eqref{eq:anchor_originated_edges}).
    
    \Statex 
    (b) Deduce candidate pairs of anchored \textit{triangles} $\mathcal{T}$ (Eq. \eqref{eq:anchor_originated_triangles}).
    
    \Statex 
    (c) Filtered  $\mathcal T' \subseteq \mathcal T$ based on \textit{opposite-edge} lengths (Eq. \eqref{eq:filtered_anchor_originated_triangles}).
    
    \Statex 
    (d) Sort $\mathcal{T}'$ by ascending \textit{triangle distortion} $d_{\Delta}$ (Eq.~\eqref{eq:triangle_length_distortion}).

\vspace{1mm}
\State \textbf{Verify} \textit{(Evaluate transforms on candidate triangle pairs)}

\Statex \textbf{for} $n=1,\ldots,N_{\mathrm{iter}}$ \textbf{do}

    \Statex \;\;(a) Compute transform
    $(\hat{\mat{R}}_n, \hat{\vct{t}}_n)$ for $\tau_n\in\mathcal T'$
    via $\operatorname{Kabsch}$ \eqref{eq:kabsch_rmsd}

    \Statex \;\;(b) Apply transform to $\mathcal{X}$ : 
    $\hat{\mathcal X}_n = \hat{\mat{R}}_n\mathcal{X} + \hat{\vct{t}}_n$

    \Statex \;\;(c) Compute pairwise cost matrix
    $\mat{C}_{\varepsilon}(\hat{\mathcal X}_n,\mathcal Y)$ (Eq.~\eqref{eq:cost_matrix})

    \Statex \;\;(d) Solve LAP on $\mat{C}_{\epsilon}$ to obtain matched set $\mathcal M_n\subset\mathcal{X}\times\mathcal{Y}$

\Statex \textbf{end for}

\vspace{1mm}
\State
\textbf{Select}
    \Statex (a) Pick $\mathcal M^*=\mathcal M_n$ such that $|\mathcal M_n|$ is maximal, using smallest total residual $r(\mathcal M_n)$ (Eq. \eqref{eq:mean_residual}) to break ties
    \Statex (b) $(\mat{R}^*,\vct{t}^*)=\operatorname{Kabsch}(\mathcal M^*)$ (Eq.~\eqref{eq:kabsch_rmsd})

\end{algorithmic}
\end{algorithm}
The above-reviewed methods suffer in our setting, due to the internal symmetries and the presence of spurious and noisy estimates in image-source point clouds. To fight this, we note that the position of the \textit{true source} is easily identified and localized from RIR signals (first echo) and exploit it as an \textit{anchor point}, denoted $\vct x_0\in\mathcal{X}$ and $\vct y_0\in\mathcal{Y}$ in each set.
We propose \textbf{ARCTIC} (\textbf{A}nchor-based \textbf{R}egistration using \textbf{C}ongruent \textbf{T}r\textbf{I}angle \textbf{C}onsensus), a new \textit{hypothesize-and-verify} algorithm, summarized in Algorithm~\ref{alg:arctic} and detailed below. The complete pipeline is illustrated in Fig.~ \ref{fig:proposed_pipeline}. Code to reproduce the method and results is available on GitHub\footnote{\url{https://github.com/tguffroy/ARCTIC}}.

\vspace{1mm}
\noindent \textbf{Hypothesize.}
First, a list of candidate \textit{congruent-triangle} pairs is built. Naively enumerating all triangle pairs without filtering has an asymptotic complexity of $O(N^3M^3)$ which is prohibitive in our settings (billions of checks).
Instead, we impose the anchor to be one of the vertices of every triangle, and filter them by their edge-length consistency. We start by building a list of candidate pairs of anchor-originated \textit{edges}:
%
\begin{equation}
    \label{eq:anchor_originated_edges}
    \hspace{-1mm}
    \mathcal E \hspace{-1mm}=\hspace{-1mm}
    \Bigl\{
    \hspace{-0.5mm}
    (\vct{x},\vct{y}) \hspace{-1mm}\in\hspace{-1mm} \mathcal{X}_{\setminus \{\vct x_0\}} \hspace{-0.5mm}\times\hspace{-0mm} \mathcal{Y}_{\setminus\{\vct y_0\}};
    \big|\|\vct{x}-\vct{x}_0\|-\|\vct{y}-\vct{y}_0\|\big|
    \hspace{-1mm}<\hspace{-1mm}
    \varepsilon
    \Bigr\},
\end{equation}
%
where the threshold $\varepsilon$ is a hyperparameter representing \textit{geometrical consistency} (in meters) throughout the method. $\mathcal E$ contains pairs of points whose distance to the true source is consistent. We deduce a set of anchored \textit{triangles} from $\mathcal E$:
\begin{equation}
    \label{eq:anchor_originated_triangles}
    \hspace{-2mm}\mathcal T \hspace{-1mm}=\hspace{-1mm} \Bigl\{
    \hspace{-1mm}
    \big((\vct{x}_0,\vct{x}_i,\vct{x}_j),(\vct{y}_0,\vct{y}_k,\vct{y}_l)\big);
    (\vct{x}_i,\vct{y}_k),\hspace{-0.5mm}
    (\vct{x}_j,\vct{y}_l)
    \hspace{-0.5mm}\in\hspace{-0.5mm}
    \mathcal{E}
    \Bigr\}.
\end{equation}
We filter $\mathcal T$ by keeping only triangle pairs whose opposite-edge length is consistent:
\begin{equation}
    \label{eq:filtered_anchor_originated_triangles}
    \hspace{-1mm}\mathcal T' \hspace{-1mm}=\hspace{-1mm} \Bigl\{
    \hspace{-1mm}
    \left(
    \begin{array}{l}
    \hspace{-2mm}(\vct{x}_0,\vct{x}_i,\vct{x}_j), \\
    (\vct{y}_0,\vct{y}_k,\vct{y}_l)
    \end{array}
    \hspace{-2mm}
    \right)\hspace{-1mm}\in\hspace{-1mm}\mathcal{T};
    \big|\|\vct{x}_i-\vct{x}_j\|
    -
    \|\vct{y}_k-\vct{y}_l\|\big|
    \hspace{-1mm}<\hspace{-1mm}
    \varepsilon
    \Bigr\}.
\end{equation}
Finally, we introduce a \textit{triangle-distortion measure} $d_\Delta$. For a triangle pair $\bigr((\vct{x}_0,\vct{x}_i,\vct{x}_j),(\vct{y}_0,\vct{y}_k,\vct{y}_l)\big)\in\mathcal{T}'$, we define $d_\Delta$ by the root mean square of the edge-length differences:
\begin{equation}
    \label{eq:triangle_length_distortion}
    d_\Delta = \sqrt{(e_{i,0,k,0}^2+e_{j,0,l,0}^2+e_{i,j,k,l}^2)/3},
\end{equation}
where $e_{i,j,k,l} = \big| \|\vct x_i - \vct x_j\| - \|\vct y_k - \vct y_l\| \big|$ for all $(i,j,k,l)$. $\mathcal T'$ is then sorted by ascending order of $d_\Delta$, so that the most edge-length-consistent triangle pairs are checked first.

\vspace{1mm}
\noindent \textbf{Verify.}
We iterate over the first $N_\text{iter}$ triangle pairs of $\mathcal T'$. Let us consider the $n$-th triangle pair ($n < N_\text{iter}$). Using the Kabsch algorithm \eqref{eq:kabsch_rmsd}, we compute the corresponding hypothesis transformation $(\hat{\mat R}_n, \hat{\vct t}_n)$ and apply it to $\mathcal X$ to obtain the transformed point cloud $\hat{\mathcal X}_n = \hat{\mat{R}}_n\mathcal{X} + \hat{\vct{t}}_n$. Next, we solve a \textit{linear assignment problem} (LAP) to find a one-to-one match set $\mathcal M_n\subseteq\mathcal{X}\times\mathcal{Y}$ between clouds. To do so, we define a cost matrix $\mat C_\varepsilon(\hat{\mathcal X}_n, \mathcal Y) \in \mathbb{R}^{|\mathcal X|\times |\mathcal Y|}$ whose entries are defined by:
%
%
\begin{equation}
\label{eq:cost_matrix}
    \hspace{-1mm}c^{\epsilon}_{i,j} \hspace{-1mm} = \hspace{-1mm}
    \begin{cases}
        \hspace{-1mm}\| \hat{\vct x}_{n,i} - \vct  y_j \|
        \;\text{if } 
        e_{i,0,j,0} \leq  \varepsilon \;\text{and}\; \| \hat{\vct x}_{n,i} - \vct  y_j \| \hspace{-0.5mm}< \hspace{-0.5mm}\varepsilon\\
        +\infty \;\; \text{otherwise}.
    \end{cases}\hspace{-2mm}
\end{equation}
By construction, this removes matches with alignment errors above the prescribed $\varepsilon$.
The LAP is solved using Scipy's \texttt{\small linear\_sum\_assignment} implementation, returning a matched set $\mathcal M_n$ of size $\leq \min(N, M)$ that minimizes the total cost $\sum_{i,j} c^{\epsilon}_{i,j}$. 
We compute the mean residual as:
\begin{equation}
    \label{eq:mean_residual}
    r(\mathcal M_n) = \frac{1}{|\mathcal M_n|}\textstyle\sum_{(\vct x, \vct y) \in \mathcal M_n} \| \hat{\mat R}_n \vct x + \hat{\vct t}_n - \vct y \|.
\end{equation}

\vspace{-1mm}
\noindent \textbf{Select.}
We select the set of matched points $\mathcal M^* = \mathcal M_n$ with largest size $|\mathcal M_n|$, using $r(\mathcal M_n)$ to break ties. The final transformation $(\mat R^*, \vct t^*)$ is obtained by a final run of the Kabsch algorithm on the full set $\mathcal M^*$.

\vspace{-3mm}
\section{Experiments and Results}
\vspace{-3mm}
\subsection{Dataset}
\vspace{-1mm}
\label{sec:data}
ARCTIC is evaluated on both simulated and real data. The former consist of 1000 shoebox rooms of dimensions randomly drawn between $3\hspace{-0.5mm}\times\hspace{-0.5mm}3\hspace{-0.5mm}\times\hspace{-0.5mm}2.1$ and $7\hspace{-0.5mm}\times\hspace{-0.5mm}7\hspace{-0.5mm}\times\hspace{-0.5mm}5$~m. A sound source and two microphone arrays are randomly placed inside each room, with a minimum distance to reflectors of 0.5~m. The arrays are both at least 1~m away from the source, and their orientations are uniformly randomized. For each scene, we use \texttt{pyroomacoustics} \cite{scheibler18} to generate the corresponding image source (IS) point cloud, with the ISM up to order 4. The point-cloud coordinates in each array's frame are calculated to obtain $\mathcal{X}$ and $\mathcal{Y}$. Synthetic degradations (Gaussian noise on IS positions, random IS removal) proportional to the distance to the true source are used to emulate real data, while the true source is kept untouched. IS further than 10~m from each array are removed. 500 room configurations are used to tune ARCTIC's hyperparameters via grid search, while the 500 remaining ones are used to test ARCTIC on unseen data. We retain $\varepsilon = 0.8$~m and $N_\text{iter} = 1000$ in what follows.

Real data are extracted from the recently published ARPEGE dataset \cite{pascal2026arpege}. 36 pairs of 32-channel RIRs measured with the em32 Eigenmike and a Genelec 8020 loudspeaker in different positions and orientations across 6 rooms are selected. To obtain point clouds from RIRs, we first pre-process them using \textit{Real2Sim} (\cite{prieto2026bridging}, see Sec.~\ref{subsec:RIR_Background}) retrained on a new dataset of 512-sample simulated RIRs with randomized source orientation. The pre-processed RIRs are then fed to the algorithm of \cite{sprunck22}, yielding point clouds of up to 70 IS.

\vspace{-3.5mm}
\subsection{Metrics and Baselines}
\vspace{-1.5mm}
Let us denote $(\mat R^{\textrm{gt}}, \vct t^{\textrm{gt}})$ the true transformation (ground truth), and $(\mat R^*, \vct t^*)$ the transformation estimated by ARCTIC. For evaluation, we use the same metrics as in \cite{redondi2009geometric} and \cite{valente2010self}, namely, the angular error  $\epsilon_{\vct t}$ between $\vct t^{\textrm{gt}}$ and $\vct t^*$ and the geodesic distance $\epsilon_{\mat R}$ between $\mat R^{\textrm{gt}}$ and $\mat R^*$. We also introduce the \textit{inter-array distance error} $\Delta r = \big| \|\vct t^*\| - \|\vct t^{\text{gt}} \| \big|$, and the \textit{Euclidean distance error}, given by $\Delta \vct t = \|\vct t^* - \vct t^{\text{gt}} \|$. The latter captures the relative positional error between the two arrays and accumulates the three previous ones.
We consider an estimate a \textit{success} if $\epsilon_\mat R < 15$° and $\Delta \vct t < 50$~cm. We compare ARCTIC with Super4PCS \cite{mellado2014super}, a computationally more efficient version of 4PCS (Sec.~\ref{sec:point_cloud_registration}), using its OpenGR implementation\footnote{\url{https://storm-irit.github.io/OpenGR/}}. We also compare ARCTIC to Super4PCS followed by 500 ICP iterations (\cite{besl92}, Sec.~\ref{sec:point_cloud_registration}) for refinement (4PCS+ICP). Based on a grid search, the hyperparameters used for Super4PCS are \texttt{\small overlap} = 0.5 and $\delta$ = 1.3~m on simulated data; \texttt{\small overlap} = 0.4 and $\delta$ = 1.4~m on real data.

\vspace{-3.5mm}
\subsection{Results}
\vspace{-1.5mm}
\begin{table}[t!]
    \centering
    \small
    \setlength{\tabcolsep}{1.7pt}
    \begin{tabular}{|c|c|c|c c c c|}
    \hline
    \textbf{Data} & \textbf{Method} & \textbf{Success} & $\bar \epsilon_{\mat R}$(°) & $\bar \epsilon_{\vct t}$(°) & $\Delta \bar r$ (cm) & $\Delta \bar{\vct t}$ (cm) \\
    \hline \hline

    \quad & 4PCS & 81 (16.2\%) & 4.58 & 10.06 & 15.0 & 34.4 \\
    Sim. & 4PCS+ICP & 135 (27.0\%) & \textbf{2.48} & 6.50 & 11.2 & 22.8 \\
    \quad & \textit{ARCTIC} & \textbf{481} \textbf{(96.2\%)} & \textbf{2.49} & \textbf{4.06} & \textbf{3.8} & \textbf{8.8} \\
    \hline \hline

    \quad & 4PCS & 8 (22.2\%) & 6.15 & 6.62 & 12.2 & 23.8 \\
    Real & 4PCS+ICP & 23 (63.9\%) & \textbf{4.71} & 5.61 & 10.7 & 20.6 \\
    \quad & \textit{ARCTIC} & \textbf{36} \textbf{(100\%)} & \textbf{4.82} & \textbf{4.81} & \textbf{9.5} & \textbf{19.4} \\
    \hline

    \end{tabular}
    \caption{Comparison of ARCTIC (proposed) with Super4PCS \cite{mellado2014super} and Super4PCS+ICP \cite{besl92} on simulated and real data. The metrics are averaged over successes only.
    \vspace{-5mm}}
    \label{table:super4pcs_comparison}
\end{table}
Table \ref{table:super4pcs_comparison} summarizes the results. Our best baseline 4PCS+ICP only reaches 27\% and 64\% successful alignments on simulated and real data respectively, highlighting the difficulty of the task. Around 75\% of these incorrect alignments can be attributed to symmetries and repeated patterns naturally occurring in the IS point clouds. Indeed, we detected \textit{flipped} alignments ($\epsilon_R > 165^{\circ}$) on 55\% of simulated data and 27.8\% of real data. For successful alignments, the metrics are significantly improved by using the ICP refinement. Meanwhile, ARCTIC strongly outperforms the two baselines on both test sets, achieving a near 100\% success rate while equaling or decreasing alignment errors across all metrics. These encouraging results can be partly attributed to the use of the true source as a natural anchor, which the baselines do not exploit.
Quantitatively, angular errors below $5^{\circ}$ and inter-array distance errors below $10$~cm are achieved on real data, showing potential for distributed array processing applications down the line. Errors are nearly twice as low on simulated data, which may be partly explained by model mismatch and acquisition noise, and partly by small annotation errors in the ARPEGE dataset. Indeed, the localization method underlying our approach is expected to  estimate the \textit{acoustic center} of sources, which may be distinct from the annotated centers, and is even frequency-dependent for most loudspeakers \cite{brandner2018dirpat}. 




\begin{figure}[t!]
  \includegraphics[width=0.5\textwidth]{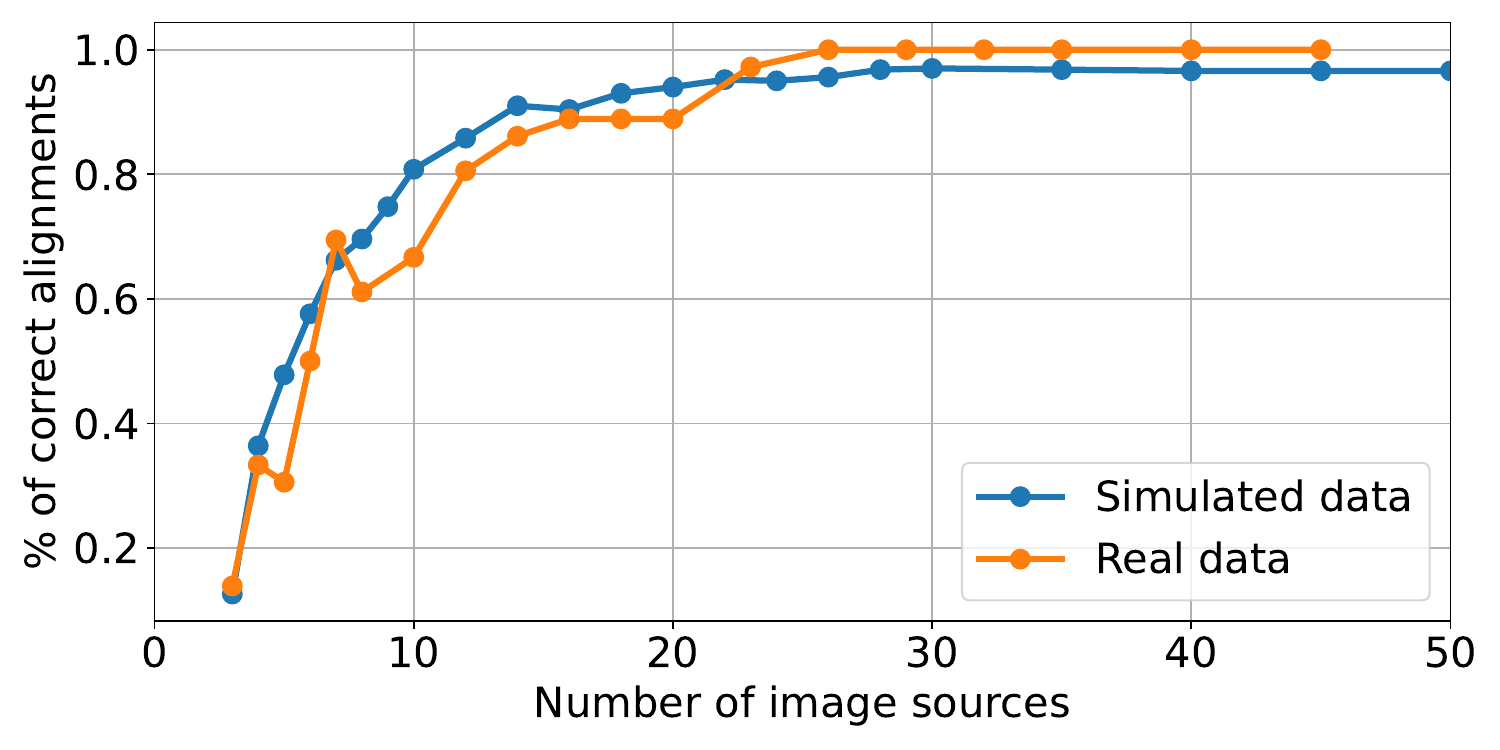}
  \vspace{-8mm}
  \caption{Influence of the number of image sources on the proportion of correct alignments using ARCTIC.
  \vspace{-5mm}}
  \label{fig:is_influence}
\end{figure}

Finally, we investigate the influence of the number of IS on ARCTIC's performance. IS are sorted by ascending order of distance to the microphone array, and ARCTIC is run on a progressively higher number of nearest IS. Fig.~\ref{fig:is_influence} shows success rates as a function of this number on both simulated and real data. Interestingly, performance steadily increases by using up to 25 IS, highlighting the under-exploited potential of leveraging higher-order acoustic reflections in microphone array processing. In particular, we observed that sources of order up to 4 caused by floor-ceiling couplings were successfully exploited by the method on real data. Performance plateaus after this point, owing to the lesser reliability of IS localization as echo arrivals near the reverberation mixing time.

\vspace{-3mm}
\section{Conclusion and Future Work}
\vspace{-2mm}
We introduced ARCTIC, an initialization-free algorithm for registering image source point clouds, and used it to self-localize two microphone arrays from a single source in a reverberant environment. By exploiting the source as an anchor, ARCTIC mitigates geometric ambiguities inherent to the task, sharply improving robustness compared to two baselines, unlocking consistent and accurate self-localization on real data. Future work will include replacing input RIRs by less-controlled signals such as hand claps. This adds one or two degrees of freedom to the problem, since the source's emission time in each array's clock is then unknown. A promising lead is to infer it by exploiting geometrical priors on image-source point clouds. A further extension is to account for dynamic scenarios with moving arrays or sources.



\end{document}